\documentclass[prl,twocolumn,showpacs,preprintnumbers,amsmath,amssymb,floatfix]{revtex4}

\usepackage{graphicx}
\usepackage{dcolumn}
\usepackage{bm}
\usepackage{color}
\usepackage{url}
\usepackage{amsmath}

\begin{document}
\newcommand{\Cardiff}{School of Physics and Astronomy, Cardiff University, Queens Building, CF24 3AA, Cardiff, United Kingdom}

\title{
Is black-hole ringdown a memory of its progenitor? 
}

\author{Ioannis Kamaretsos}\author{ Mark Hannam}\author{B.S.~Sathyaprakash}
\affiliation{\Cardiff}

\begin{abstract}~
We have performed an extensive numerical study of coalescing black-hole binaries to 
understand the gravitational-wave spectrum of quasi-normal modes excited in the merged 
black hole. Remarkably, we find that the masses and spins of the progenitor are clearly 
encoded in the mode spectrum of the ringdown signal. Some of the mode amplitudes carry 
the signature of the binary's mass ratio, while others depend critically on the spins. 
Simulations of precessing binaries suggest that our results carry over to \emph{generic} 
systems. Using Bayesian inference, we demonstrate that it is possible to accurately measure 
the mass ratio and a proper combination of spins even when the binary is itself invisible 
to a detector. Using a mapping of the binary masses and spins to the final black 
hole spin, allows us to further extract the spin components of the progenitor. Our results could 
have tremendous implications for gravitational astronomy by facilitating novel tests of general 
relativity using merging black holes.
\end{abstract}
\maketitle

\paragraph{Introduction.---}

A black-hole-binary merger produces a single black hole that quickly ``rings down'' to the 
Kerr solution, fully characterised by its mass and angular momentum. 
It is well known that the frequencies and damping times of the ringdown gravitational 
waves (GWs) are described by the same two parameters (see, e.g., Ref.~\cite{1980ApJ...239..292D} 
and references therein). However, the mode distribution of the ringdown \emph{amplitudes} 
depends on the progenitor. Recently Kamaretsos {\it et.al.}~\cite{Kamaretsos:2011um} 
suggested that we could exploit this fact to measure properties of the 
progenitor from the ringdown signal. 
This was demonstrated by using a set of numerical-relativity simulations of non-spinning 
binaries parametrized by the mass ratio and constructing a signal model reflecting the clear 
mass-ratio dependence of the ringdown mode amplitudes.

It follows that in general the ringdown amplitudes will depend on all 
eight binary parameters (the two masses, plus the vector components of 
each black hole's spin). In this Letter we report two remarkable results. 
First, that at least some of the mode amplitudes depend \emph{only} 
on the mass ratio of the progenitor binary, \emph{largely independent of the spins.} 
Therefore, we should be able to use the
ringdown to measure the individual masses of a binary even when we cannot 
observe the binary itself!  Second, one other mode amplitude carries a clear 
signature of the \emph{spins} of the progenitor black holes and depends 
on an effective spin parameter related to the difference in the spin 
magnitudes. In the case of aligned spins ({\em i.e.}, non-precessing binaries), 
this fact, along with a mapping of the progenitor configuration to the final black hole 
spin ~\cite{Barausse:2009uz} allows us to determine the individual spin components from the ringdown phase alone. 

We show that progenitor parameters \emph{can} be measured with good accuracy
with the Einstein Telescope (ET)~\cite{DSD}. 
If mode amplitudes can be extracted from GW observations, they could be 
used to test strong-field general relativity, study the nature of the 
merged object, especially if it is a naked singularity, and
as the only means to observe the formation of black holes when the inspiral phase
of the signal is outside a detector's sensitivity band. 

The physical origin of the mode-amplitude relations is unclear;
but we note a relation to post-Newtonian inspiral results, 
raising questions for future research.

\paragraph{Background.---} 
For a black hole of mass $M,$ located at a distance $D,$ the plus and cross 
polarizations, $h_+$ and $h_\times,$ of GWs emitted  
due to quasi-normal mode oscillations can be written to a good approximation as 
\begin{eqnarray*}
h_{+}(t) = +\frac{M}{D}\sum_{\ell ,m} A_{\ell m}\, Y^{\ell m}_{+}\,
e^{-t/\tau_{\ell m}} \cos(\omega_{\ell m}t - m\phi + \varphi_{\ell m}), \label{eq:hplus}\\
h_{\times}(t) = -\frac{M}{D}\sum_{\ell ,m} A_{\ell m}\, 
Y^{\ell m}_{\times}\, e^{-t/\tau_{\ell m}} \sin(\omega_{\ell m}t-m\phi + \varphi_{\ell m}),
\label{eq:hcross}
\end{eqnarray*}
for $t\ge 0,$ where only the first (least damped) overtone is kept and 
the rest are omitted.
Here $A_{\ell m},$ $\omega_{\ell m},$ $\tau_{\ell m}$ are the 
mode amplitudes, frequencies and damping times, respectively, 
$Y^{\ell m}_{+,\times}(\iota)$ are related to $-2$ spin-weighted 
spherical harmonics that depend only on the inclination $\iota$ 
of the black hole's spin axis to the observer's line-of-sight 
\cite{Berti:2007a}, $\phi$ is the azimuth angle at which the black 
hole is observed with respect to a suitably chosen frame and $\varphi_{\ell m}$
the initial phase angles of the modes.

Black hole perturbation theory can be used to compute the mode 
frequencies and damping times~\cite{BCW05}, but \emph{not}
the mode amplitudes 
$A_{\ell m}$, which depend on the nature of the perturbation---in our 
case, a highly distorted black hole that results from the merger.  
Instead, we must use numerical simulations to calculate the 
mode spectrum and its dependence on the 
progenitor parameters \cite{Berti:2007b,Pan:2011gk,Kamaretsos:2011um}.

\begin{figure*}
\centering
\includegraphics[width=0.45\textwidth]{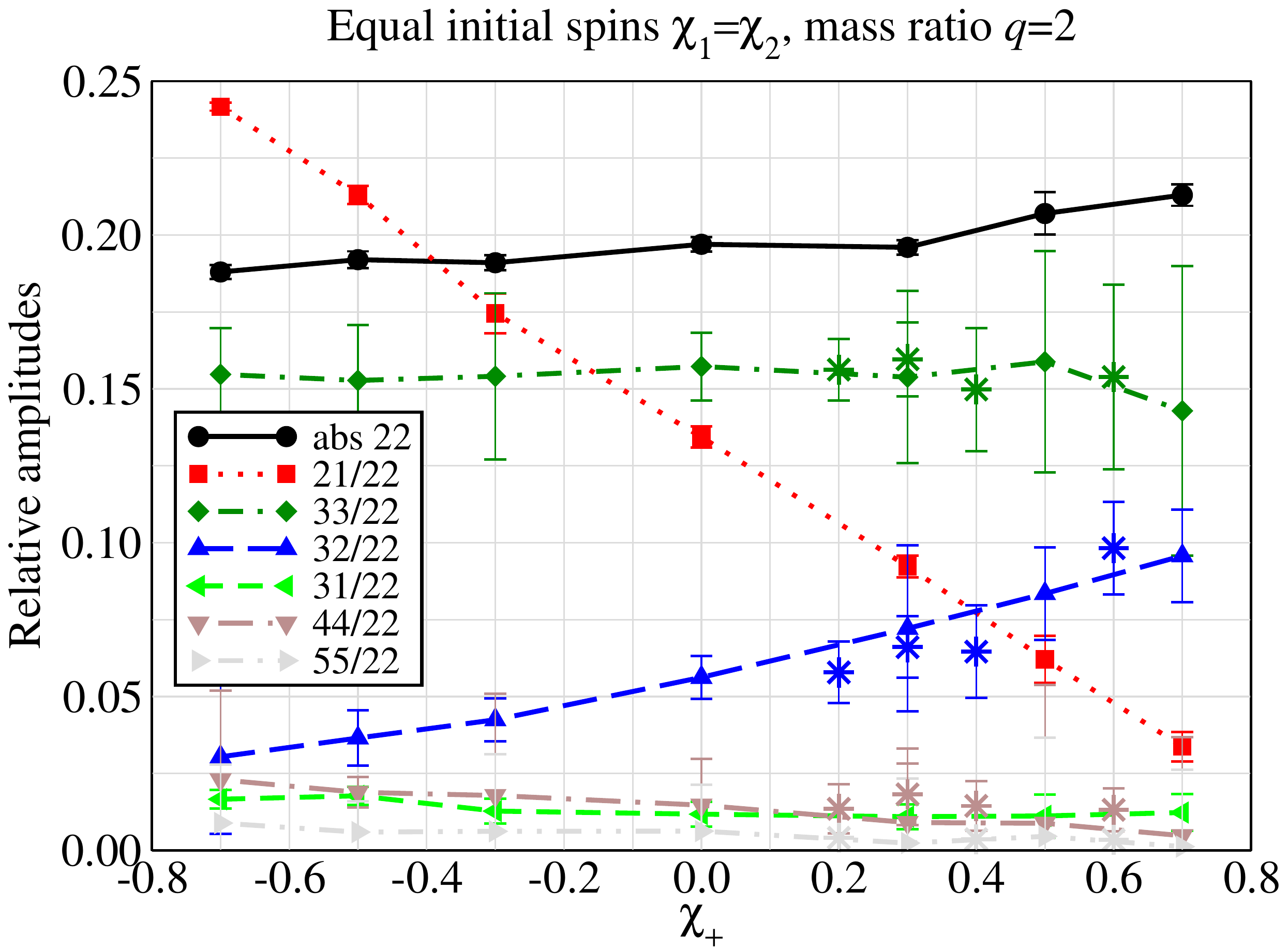}
\includegraphics[width=0.45\textwidth]{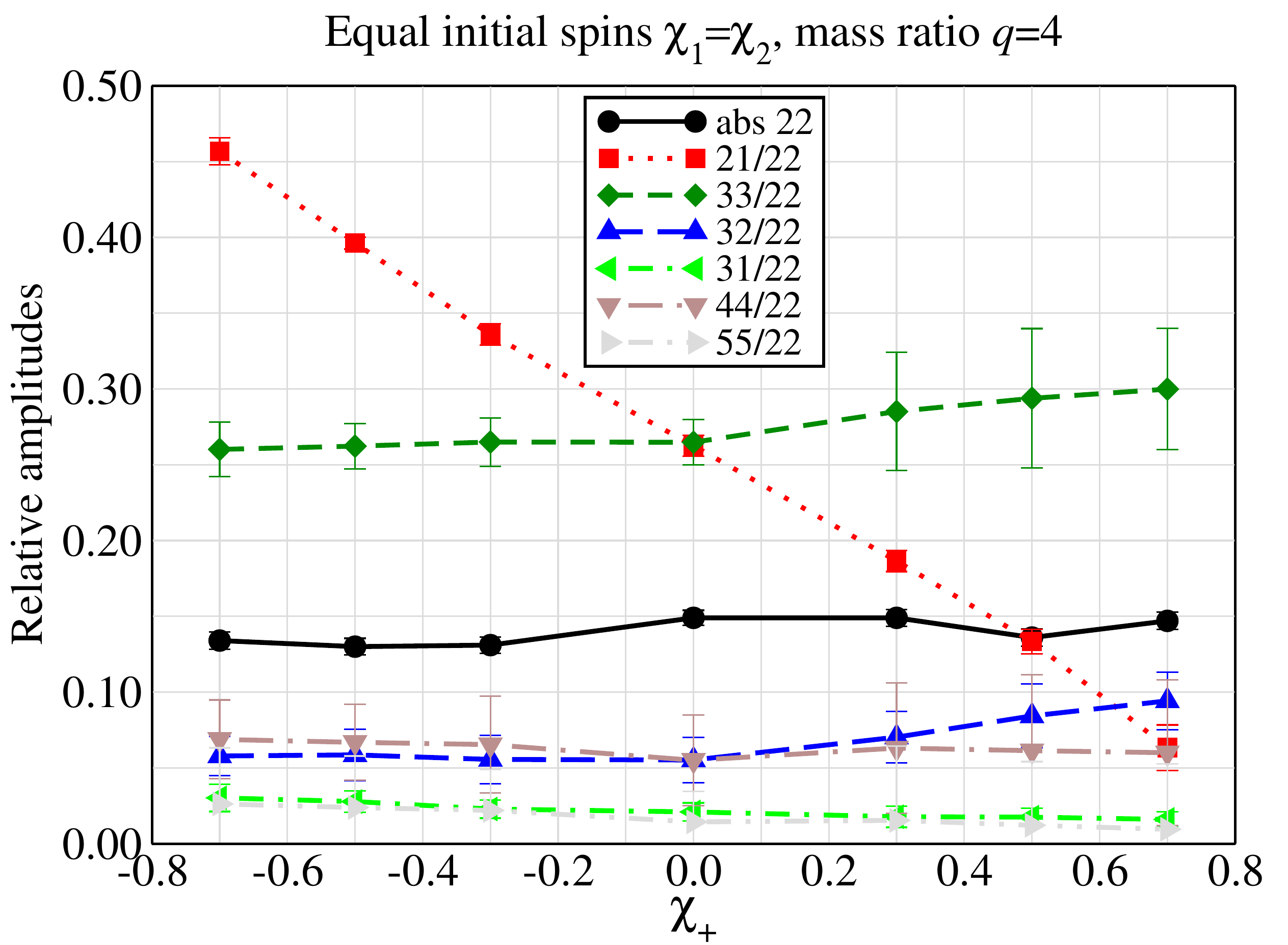}
\caption{Quasi-normal mode amplitudes of binaries with aligned spins and 
mass ratio $q=2$ (or $\nu=2/9,$ left panel) and $q=4$ (or $\nu=4/25,$ right panel). The values from 
the non-spinning binary simulations are at $\chi_+=0.$ Also shown in the left panel, with asterisks, are the results 
from the $q=2$ equal initial $\chi_i$ precessing simulations. Note that for the 22 mode, the absolute amplitudes are 
always shown, scaled according to the final black hole mass, that is $(r/M)h_{22}.$}
\label{fig:fig1}
\end{figure*}

\paragraph{Numerical results.---}

We explored the effect of spins with a large number of numerical 
binary simulations that consisted of 2-4 inspiral orbits before merger. 
There were three sets of simulations: (1) binaries with non-precessing
equal spins $\chi_i = S_i / m_i^2 = \{0, \pm0.3, \pm0.5, \pm0.7\}$ and
mass ratios $q = m_1/m_2 = \{2,4\}$, (2) systems with anti-aligned non-precessing
spins such that the final black-hole spin was the same as that for the 
corresponding non-spinning binary for $(q,\,\chi_{\rm fin}) = (2,\,0.62),$ 
$ (3,\,0.54),$ and $(4,\, 0.47),$ using the final-spin fits 
in~\cite{Barausse:2009uz,PhysRevD.78.081501} and
(3) four $q=2$ precessing binaries having equal initial spins with $(x,\, y,\, z)$ components 
equal to $(0.2,\, 0,\, 0), (0,\, 0.4,\, 0), (0.6,\, 0,\, 0)$ and $(0.2,\, 0.2,\, 0.1)$, where the orbital plane lies 
on $xy.$ There were a total of 40 configurations, not including additional tests to verify that the results 
were robust against changes in the number of inspiral orbits.

All simulations were performed with the BAM code~\cite{Brugmann:2008zz}.
As is standard,
the error bars in the amplitudes were estimated by varying the numerical resolution 
and GW extraction radius. The highest resolution near the black holes was 
$\sim m/35$, where $m$ is the mass of the smallest black hole, and the GW 
signal was typically calculated at $140\,M_{in}$ from the source. 
The ringdown amplitudes $A_{\ell m}$ were computed by 
fitting an exponential decay function to the data from $t = 10\,M$ after 
the peak of the $(2,2)$ luminosity, until the point where the signal was dominated 
by numerical noise. $A_{22}$ and $A_{21}$ are typically 
accurate to within 2\%, and $A_{33}$ and $A_{32}$ to within 10\%. The 
weaker modes are too noisy to be measured accurately, and are shown only 
for qualitative comparison.

\begin{figure*}
\centering
\includegraphics[width=0.45\textwidth]{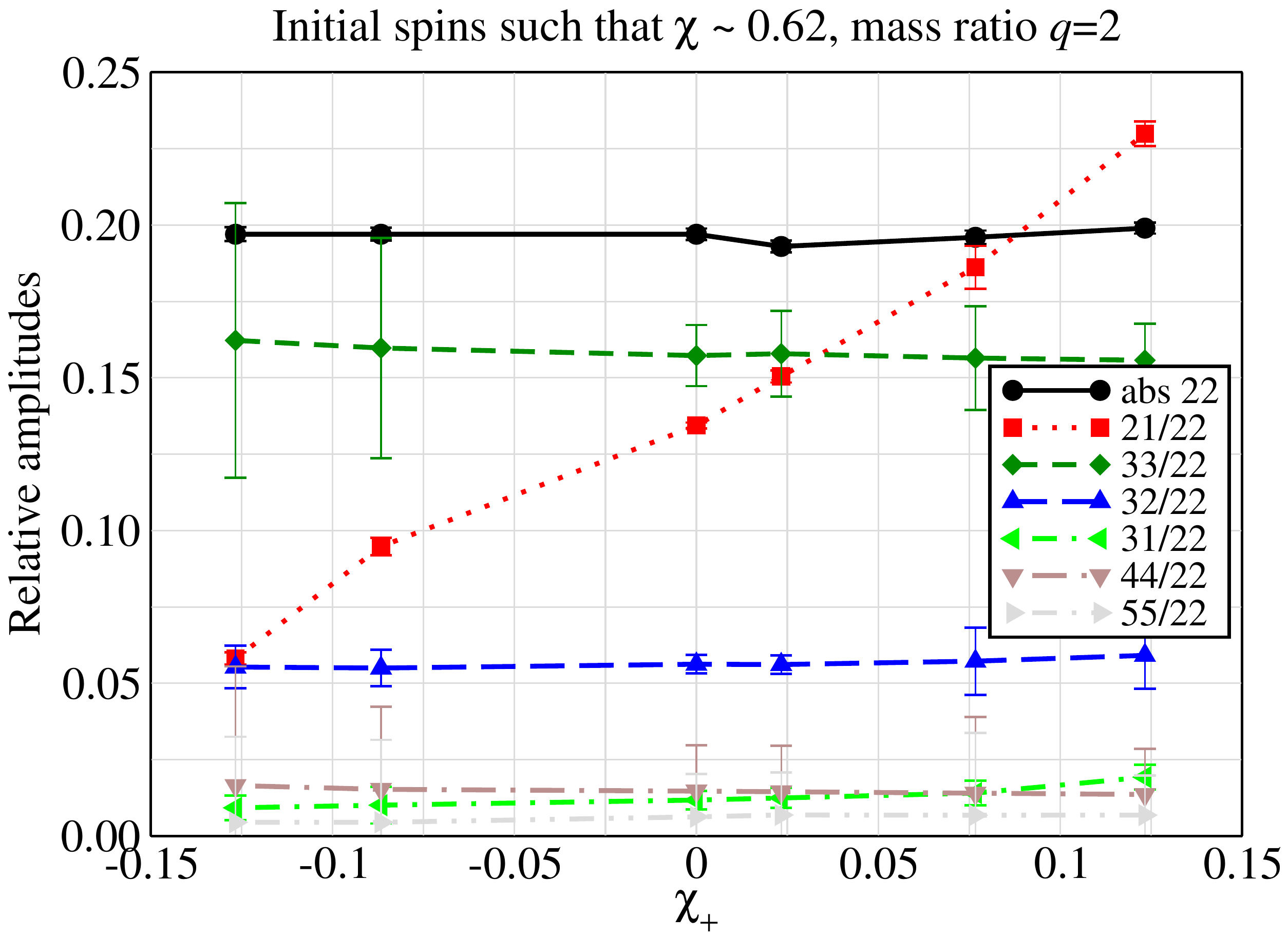}
\includegraphics[width=0.45\textwidth]{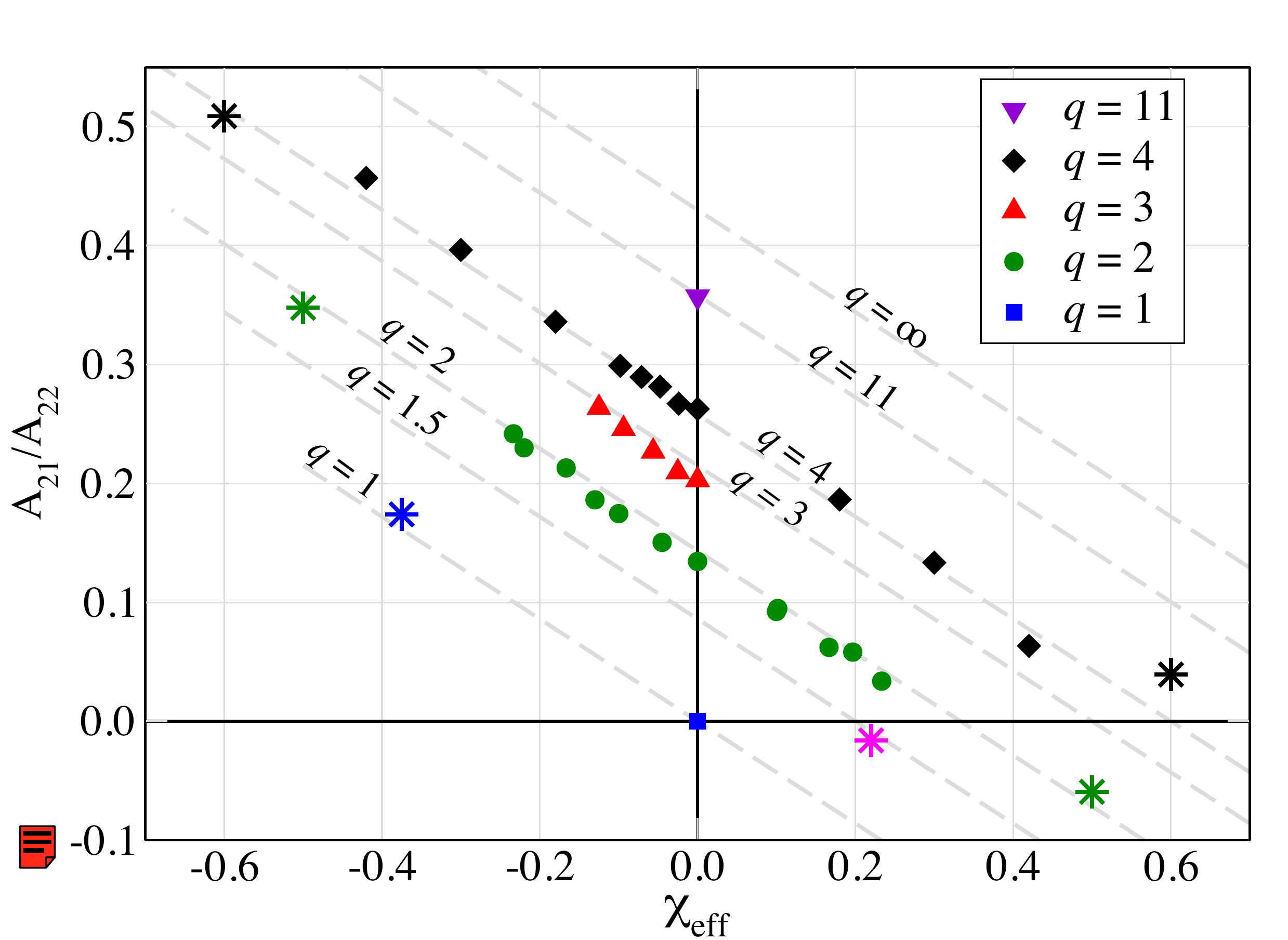}
\caption{{\em Left} panel plots the amplitude of the various modes as a function of the total spin parameter 
$\chi_+$ for the $q=2$ simulations that end in a black hole of $\chi \simeq 0.62.$ Modes 22, 
33 are again rather insensitive to progenitor spins, while 21 varies by nearly a factor of 5. {\em Right} 
panel plots the 21 amplitudes from {\em all simulation sets} as a function of an effective spin term 
$ \chi_{\rm eff}$ allowing us to estimate this parameter from a measurement. 
We verified our predictions with additional simulations marked with asterisks.
}
\label{fig:fig2}
\end{figure*}

Figure \ref{fig:fig1} shows the results for the first set of simulations, of 
equal-spin binaries. The amplitudes of the seven strongest modes ($A_{\ell m} =
A_{\ell -m}$ for non-precessing binaries) are plotted as a function of 
a total spin parameter $\chi_+=(m_1\,\chi_1+m_2\,\chi_2)/M_{in}$,
where $M_{in} = m_1 + m_2$ and $\chi_+ = \chi_i$ for these cases. This is the same 
spin parameter that has been used in recent phenomenological models of binary 
waveforms~\cite{Ajith:2009bn,Santamaria:2010yb}. The amplitudes are all 
relative to the 22 mode, for which we show the absolute amplitude.

We see immediately that $A_{22}$ and $A_{33}$ change with mass ratio,
but vary only weakly with respect to spin. In contrast, $A_{21}$ varies strongly
with spin. Figure \ref{fig:fig1}, therefore, suggests that the 22 and 33 modes
carry information about the progenitor mass ratio, and the 21 mode
carries information about the effective total spin. 

The second series of simulations tests this hypothesis. 
For each mass ratio, this set generates approximately the same final black-hole 
with different progenitor spin configurations. The goal was to show that the mode amplitudes
carried a signature of the progenitor spins independently of the final black hole spin. 
The mode amplitudes for the $q=2$ case are shown in the left panel of Fig.\ \ref{fig:fig2},
as a function of $\chi_+.$ As before, 22 and 33 show little variation, but the 21 mode changes 
by nearly a factor of five. 
This is strong evidence that the final black holes in this set are {\em not} really degenerate: 
although their mode frequencies and damping times will be identical, they will differ from one 
another in the 21 mode amplitude. 
This is consistent with studies of black-hole recoil: 
the recoil is mostly due to the interplay of the $(2,\pm2)$ and $(2,\pm1)$ 
modes~\cite{Herrmann:2007ac}, and both the
recoil and $(2,\pm1)$ mode amplitudes depend strongly on the progenitor spins.  

Unfortunately, the trend of 21 is now the opposite of that in Fig.\ \ref{fig:fig1} with
respect to $\chi_+,$ implying that the 21 mode amplitude is not determined by $\chi_+$.
Consider instead the effective spin parameter 
\begin{equation*}
\chi_{\rm eff}= \frac{1}{2}( \sqrt{1-4\,\nu}\,\chi_1 + \chi_- ),\quad
\chi_-=\frac{m_1\,\chi_1-m_2\,\chi_2}{M_{in}},
\end{equation*}
The right panel of Fig.\ \ref{fig:fig2} shows the amplitude of 21 
as a function of $\chi_{\rm eff}$ for all the simulations discussed so far. 
In {\em all} cases they are well approximated by 
\begin{equation}
\hat{A}_{21} \equiv A_{21}/A_{22} = 0.43 \left [ \sqrt{1-4\,\nu} - \chi_{\rm eff} \right ],
\label{eq:fit}
\end{equation}
which is shown by dashed lines in Fig.\,\ref{fig:fig2} for different values of $q$. 
The above equation is consistent with the expectation that $A_{21}$ will be excited 
in the case of equal mass binaries when $\chi_1 \ne \chi_2$, and also predicts that
in general it will be zero when $\chi_{\rm eff} = \sqrt{1-4\,\nu}=|m_1-m_2|/M_{in}.$ We tested these predictions 
with six additional simulations, shown in Tab.~\ref{tab:additional}.
The predicted amplitudes $\hat{A}_{21}^{\rm P}$ agree with the computed amplitudes
$\hat{A}_{21}^{\rm M}$ within error bars. 
Negative values indicate that the 21 phase is offset by $180^\circ$ with respect
to the 22 phase; in the equal-mass cases this is equivalent to swapping 
$\chi_1$ and $\chi_2$, or rotating the initial data by half an orbit.
\begin{table}[bht]
\caption{Additional simulations to test Eq.\,(\ref{eq:fit}).}
\label{tab:additional}
\begin{tabular}{c|c|c|c|c|c|c}
\hline
\hline
$q$  & 1 &   1.5 & 2   & 2   & 4   & 4   \\
\hline
$\chi_{\rm eff}$ & $-0.375$  & $0.220$  & $-0.500$  & $0.500$  & $-0.600$  & $0.600$ \\
\hline
$\hat{A}^{\rm P}_{21}$ & $0.161$ & $-0.005$ & $0.358$ & $-0.070$ & $0.516$ & $ 0.000$ \\
\hline
$\hat{A}^{\rm M}_{21}$ & 0.174 & $-0.016$ & 0.348 & $-0.059$ & 0.509  & 0.039 \\ 
\hline
\hline
\end{tabular}
\end{table}

All of these results apply to non-precessing
binaries: the progenitor spins and final spin were all parallel or anti-parallel 
to the binary's orbital angular momentum. 
This will not be true in general; the spins and orbital plane will precess 
during the inspiral, and the final
black hole's spin will be mis-aligned with respect to the pre-merger orbital 
plane. Even if the ringdown modes were rotated into an optimal frame 
by a procedure like that introduced in~\cite{Schmidt:2010it}, 
there would be an asymmetry between the $+m$ and $-m$ modes, 
since this is a signature of the out-of-plane recoil 
(see Sec.~III.A in \cite{Brugmann:2007zj}). However,
it is possible that if the ringdown modes were described in the optimal frame,
then their \emph{average} would satisfy the relations we have observed. To
test this, we simulated four precessing binaries. In each case the final spin
was mis-aligned with the initial orbital plane, but only slightly, so that to a
first approximation we could still consider the average of the $(2,\pm2)$ and
$(3,\pm3)$ modes. The results for these cases are shown in Fig.~\ref{fig:fig1}, 
and, remarkably, satisfy the same relations we have observed for 
non-precessing binaries. This provides strong evidence that our results
carry over to \emph{generic} binaries.

\paragraph{Interpretation.---}
Post-Newtonian (PN) theory provides some clue to the behavior of the 
amplitudes of the various modes. 
It is quite possible that the various modes excited during the inspiral
phase retain the memory of their structure through to the ringdown phase. 
(There are signs that this will be true from, e.g., Fig.~11 in~\cite{Baker:2008d78} for
non-spinning binaries.)
It is, therefore, instructive to look at the inspiral mode amplitudes. 
In particular, the 21 mode reads~\cite{Pan:2010hz}
\begin{equation}
h_{21} \propto \frac{\nu M_{in}v^3}{D}
\left (\sqrt{1-4\,\nu}- \frac{3}{2}\,v\,\chi_-\right ).
\end{equation}
Here $v$ is the PN expansion parameter, namely the orbital speed.
There are three points to note: Firstly, for non-spinning systems,
the 21 amplitude has {\em identical} dependence on the mass ratio 
during the inspiral and ringdown phases.  Secondly, the spin terms in the 
22 and 33 modes (indeed, all modes for which $l+m$ is even) appear at 1.5 PN order 
beyond the leading order and so spins have a negligible effect.
For $v=1/\sqrt{3},$ 22 and 33 vary by about $\sim 20\%$ when $\chi_1$ 
and $\chi_2$ change from $-0.8$ to $+0.8.$ However, for 21 
(and all odd $l+m$ modes)  the spin effect 
occurs at 0.5 PN beyond the leading order; spins 
affect odd $l+m$ modes far more strongly than they do even $l+m$ modes.
For $v=1/\sqrt{3}$, the 21 mode varies by a factor of 4.5, and the 32 mode
by 72\%, when spins vary from $-0.8$ to $+0.8.$ Finally, the dominant spin 
effect in the 21 mode amplitude is determined by the quantity 
$\chi_-.$ It is really {\em not} the total spin that determines the 
amplitude, but the difference of spins, as in the ringdown phase.

\paragraph{Measurement.---}

\begin{figure}
\centering
\includegraphics[width=0.49\columnwidth]{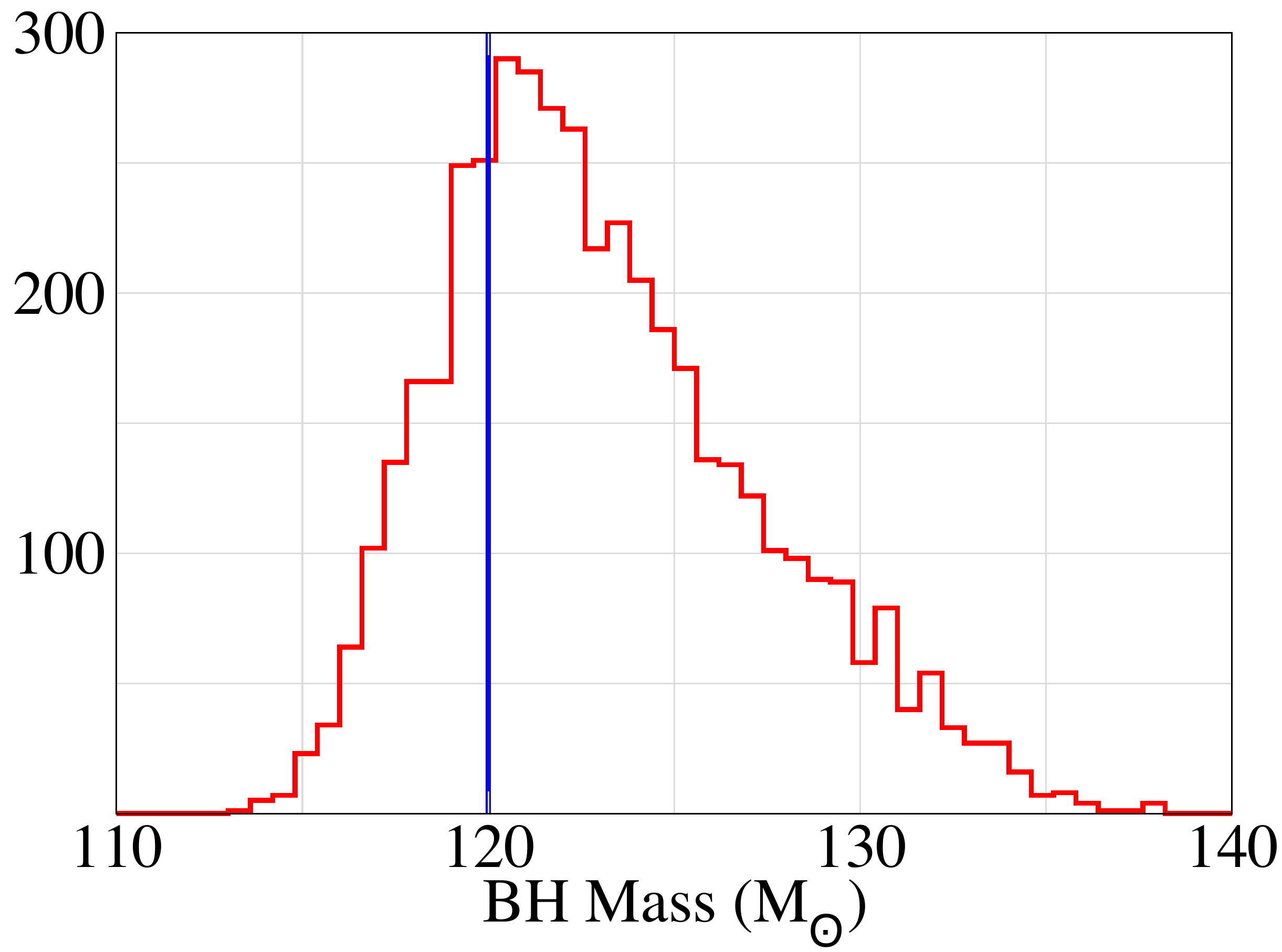}
\includegraphics[width=0.49\columnwidth]{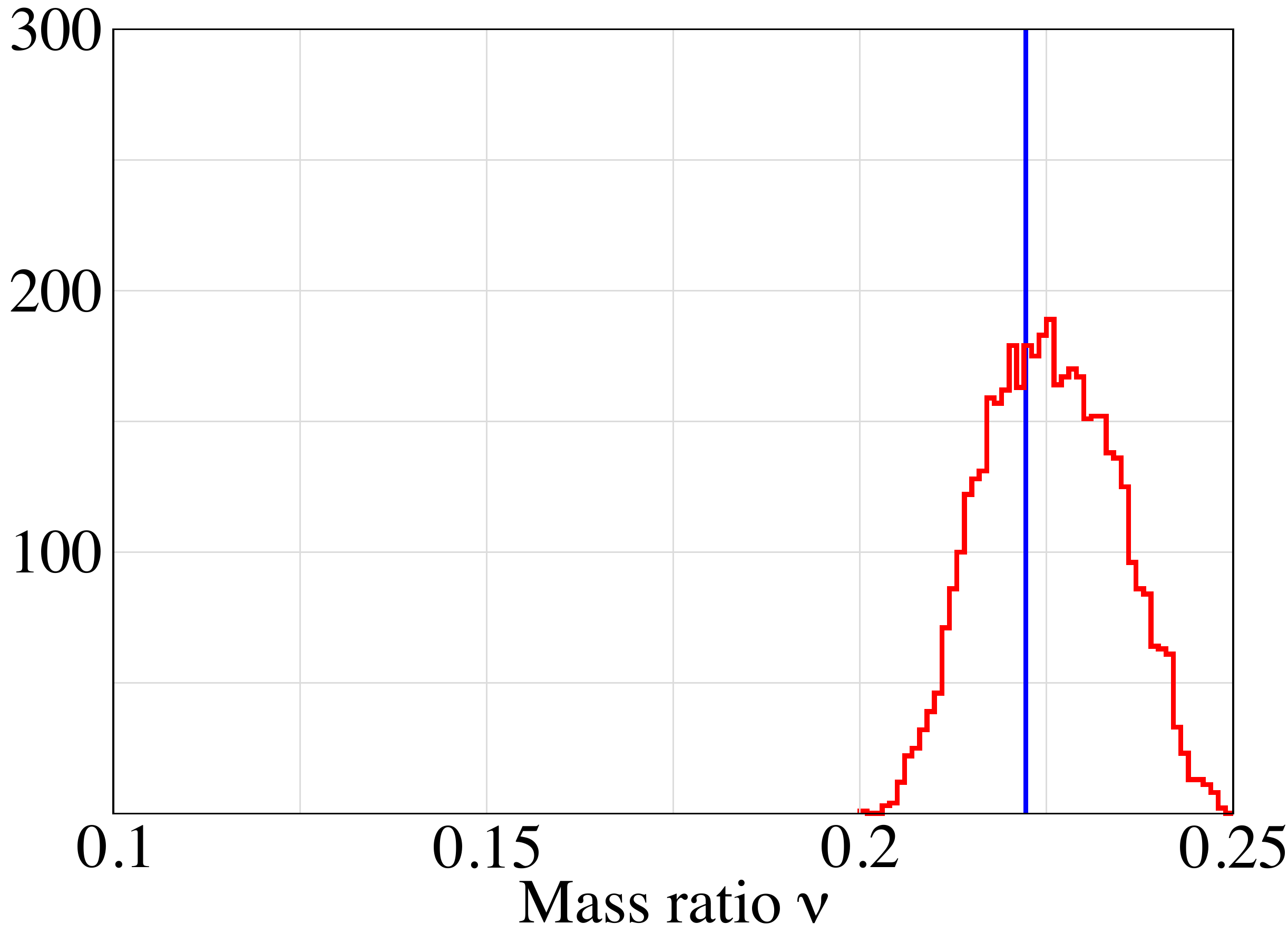}
\includegraphics[width=0.49\columnwidth]{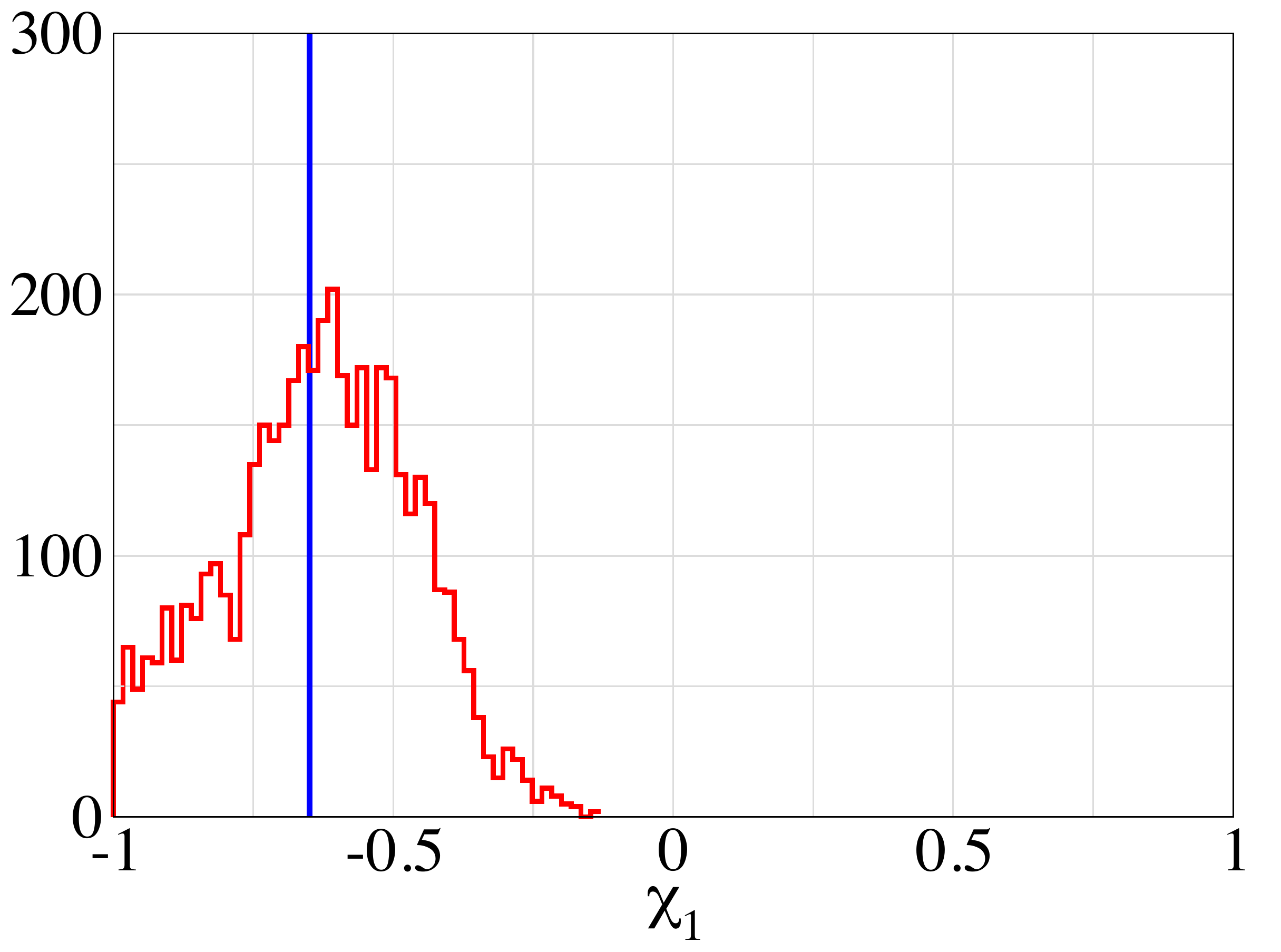}
\includegraphics[width=0.49\columnwidth]{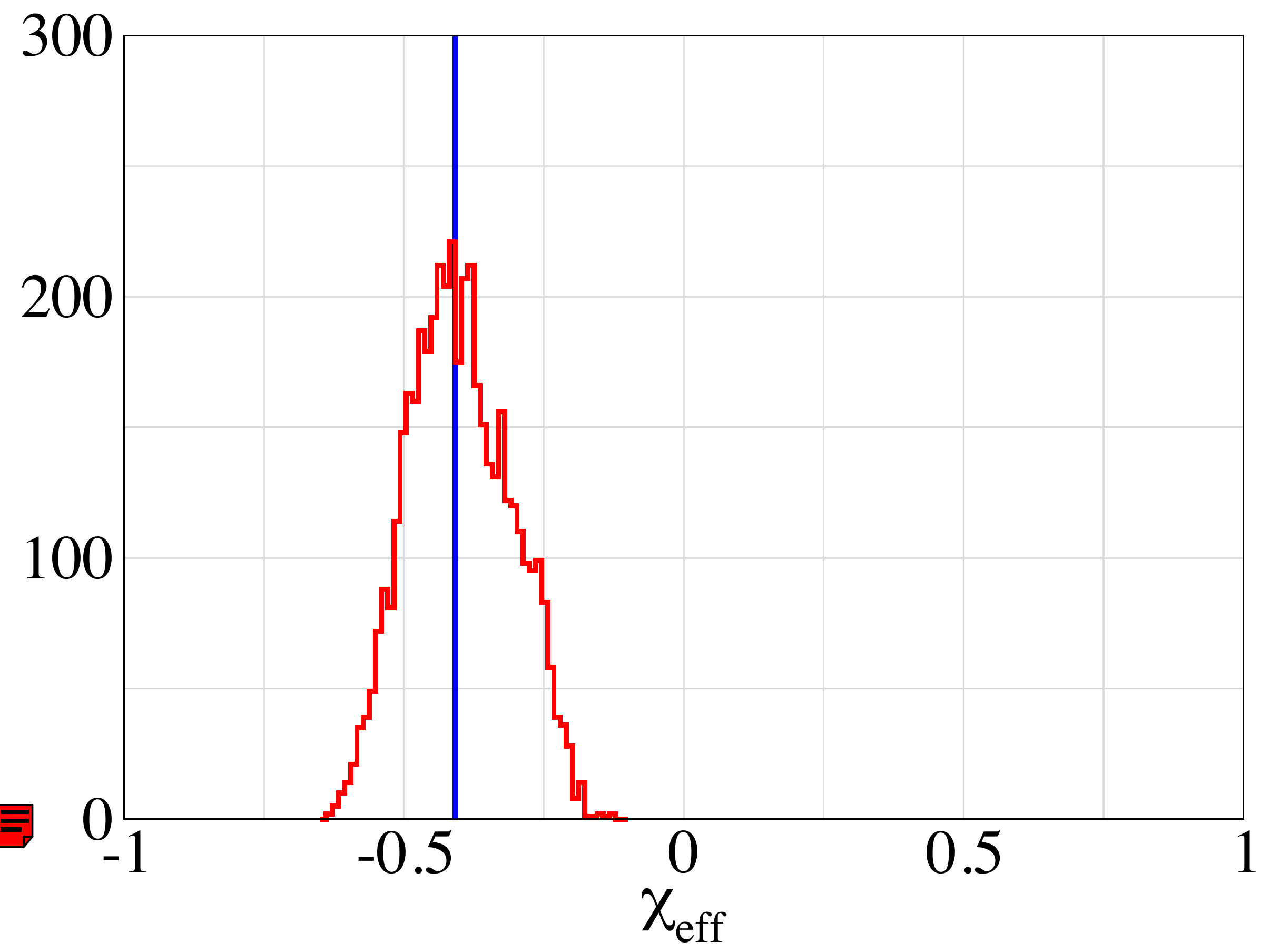}
\caption{Posterior distributions are plotted for a ringdown signal detected
with ET. The vertical lines correspond to the parameters of the signal injected
into the ET mock data stream. The source is 1\,Gpc from the detector; the SNR is $\sim 27$.}
\label{fig:fig3}
\end{figure}
To estimate how well the progenitor spins and mass ratio can be measured
we injected a ringdown signal in background noise with power spectral 
density as expected in Einstein Telescope, ET-B \cite{ET-B}, and
used Bayesian inference with nested sampling \cite{Skilling:2006,Feroz:2008} to 
detect and measure its parameters.
Our signal consists of the first three dominant modes, 22, 21 and 33,
with the 21 mode amplitude given by Eq.\,(\ref{eq:fit}) and for the 22 and 33 modes
we took $A_{22} = 0.864\,\nu,$ $\hat{A}_{33} = 0.44\,(1-4\nu)^{\,0.45}.$
The signal and the template are both is characterized by six parameters, 
$(M,\,\nu,\,\chi_1,\,\chi_2,\,D,\,t_0),$ where $t_0$ is the time-of-arrival 
of the signal at the detector.  The angles describing the location of the 
source on the sky $(\theta,\,\varphi),$ the inclination $\iota$ of the 
binary and polarization angle $\psi$, are all assumed to be known. The 
azimuth angle $\phi$ and the initial phases of the various modes 
$\varphi_{\ell m}$ are all also assumed to be zero.  These angles have 
strong correlations with the distance to the binary but not the intrinsic 
parameters. Thus, relaxing the above assumptions is not likely to have a 
big impact in the measurement of the intrinsic parameters of the source. 

The posterior distributions for four of the parameters from one of our
runs are plotted in Fig.\,\ref{fig:fig3}, which show that the parameters of
the progenitor can be quite accurately measured by using just the ringdown signal.
A more detailed study is needed to fully characterize the measurement accuracies over
the full parameter space, by incorporating other parameters such as the
sky position of the source and its inclination, assumed to be known in this work.

\paragraph{Discussion.---}
In this Letter we have addressed a question implied in Ref.\,\cite{Kamaretsos:2011um}: 
{\em can we measure the mass ratio of a generic binary from the ringdown 
signal alone?} We have found two remarkable results. First,
we \emph{can} measure the mass ratio from the ringdown 
signal, and second, we may also be able to measure the individual 
black-hole spins. In other words, both the masses and spins of the two 
component black holes could be measured purely from the rapidly 
decaying perturbation that they leave on the final merged black hole. 

The first result is demonstrated with a large numerical study of 
non-precessing binaries to show that the ratio of the amplitudes of the 
$(\ell=3, |m|=3)$ and $(\ell=2,|m|=2)$ ringdown modes carry a clear
signature of the mass ratio. Furthermore, we have evidence from a small
set of precessing-binary configurations that this signature is retained
in generic binaries. And finally, we have shown that this signature could
be accurately measured in observations with the Einstein 
Telescope. 

The second result is restricted to non-precessing binaries. 
We found that the ratio of the $(\ell=2,|m|=1)$ and $(\ell=2,|m|=2)$ mode 
amplitudes depends on a certain difference between the individual black-hole
spins. We produced a model of this spin dependence in terms of an 
effective spin parameter $\chi_{\rm eff}$, which is accurate
across a wide sampling of the non-precessing-binary parameter space. 
In a parameter-estimation exercise, where this model is injected into 
simulated Einstein Telescope noise, measurements of the final mass and
spin, and of $\chi_{\rm eff}$, can be used in conjunction with a final-spin 
fit~\cite{Barausse:2009uz,PhysRevD.78.081501} to determine the 
individual black-hole spins. 

Many questions remain open for future research. What is the physical 
origin of the observed ringdown spectrum? How do we fully model the 
ringdown signal from generic binaries? And, of most significance, what
additional astrophysics will these results allow us to learn in third 
generation GW detectors, and how precisely will we be able to test
general relativity?

\paragraph{Acknowledgements.---}

We thank S.\,Husa, T.\,Dent and J.\,Creighton for discussions. Simulations were 
performed on ARCCA Cardiff, VSC Vienna, LRZ Munich, 
MareNostrum at BSC-CNS and the 
PRACE clusters Hermit and Curie. This work was supported by
STFC UK  
grants ST/H008438/1 and ST/I001085/1.

\bibliography{ref-list}

\end{document}